\documentclass[aps,prd,twocolumn,groupedaddress,nofootinbib]{revtex4-1}
\usepackage{tensor}
\usepackage{amsmath}
\usepackage{amssymb}
\usepackage{mathtools}
\usepackage{graphicx}
\usepackage{wrapfig}
\usepackage{float}
\usepackage[
	colorlinks=true,
	linkcolor=blue,
	citecolor=blue,
	filecolor=magenta,
	urlcolor=blue
]{hyperref}

\newcommand{\rootg}{\sqrt{-g}}
\newcommand{\fRT}{f(R,T)}
\newcommand{\ft}{f_{T}(R,T)}
\newcommand{\fr}{f_{R}(R,T)}
\newcommand{\Lm}{\mathcal{L}_{m}}

\begin{document}
\title{Limits on $f(R,T)$ gravity from Earth's atmosphere}
\author{Taylor M. Ordines}
\author{Eric D. Carlson}
\email{ecarlson@wfu.edu}
\affiliation{Department of Physics, Wake Forest University, 1834 Wake Forest Road, Winston-Salem, North Carolina 27109, USA}
\date{\today}

\begin{abstract}
	We investigate changes in Earth's atmospheric models coming from the $f(R,T)$ modified theory of gravity, in which the gravitational Lagrangian is given by an arbitrary function of the Ricci scalar and the trace of the stress-energy tensor.
	We obtain a generic form for the gravitational field equations and derive the hydrostatic equation for Earth's atmosphere for leading order terms $f(R,T) = R + 2\chi T.$
	Based on the apparent accuracy of the 1976 U.S. Standard Atmosphere model, which varies no more than $10\%$ from observations, we find limits of $-1.6\times 10^{-13} \lesssim \chi \lesssim 1.8\times 10^{-13}$.
\end{abstract}

\maketitle

\section{Introduction}
	Modern cosmological observation has revealed the accelerating expansion of the universe~\cite{riess_observational_1998,perlmutter_discovery_1998,perlmutter_measurements_1999}.
	The discovery has become one of the most important developments in modern cosmology due to its apparent inconsistency with the predictions of general relativity, which state that a universe filled with a mixture of ordinary matter and radiation should experience a slowing of the expansion.
	Novel theories and modifications to general relativity have been proposed to explain the acceleration.
	Two notions in particular have been heavily investigated: either the Universe contains a great amount of dark energy, or the theory of general relativity breaks down on the cosmological scale~\cite{frieman_dark_2008}.
	
	One theory that has gained much attention for its ability to explain the expansion is $f(R)$ modified gravity~\cite{buchdahl_non-linear_1970}.
	The gravitational action, from which the Einstein equations are derived, is traditionally a linear function of the scalar curvature $R$.
	The $f(R)$ theory of gravity replaces $R$ in the gravitational action with an arbitrary function $f(R)$, with the traditional $R$ being the leading order contribution.
	Higher order gravity theories, such as the Starobinsky model~\cite{starobinsky_new_1980}, are thus generally possible.
	
	A generalization of $f(R)$ gravity proposed in~\cite{bertolami_extra_2007} incorporates an explicit coupling between the matter Lagrangian and an arbitrary function of the scalar curvature, which leads to an extra force in the geodesic equation of a perfect fluid.
	It was later shown that this extra force may account for the accelerated expansion of the universe~\cite{bertolami_accelerated_2010}.
	The inclusion of matter terms in the gravitational action was further explored in~\cite{harko_frt_2011} in $f(R,T)$ gravity, in which the gravitational Lagrangian density is an arbitrary function of both $R$ and the trace of the stress-energy tensor $T$.
	Part of the motivation of $f(R,T)$ gravity is to produce models that can act similarly to a cosmological constant without explicitly including such a constant.
	The arbitrary dependence on $T$ encapsulates the possible contributions from both nonminimal coupling and explicit $T$ terms.
		
	In general such modified theories can contain higher than first order derivatives in the Lagrangian and will thus suffer Ostrogradsky instability.
	One can in principle introduce an auxiliary field to the Lagrangian to remove the higher derivatives, resulting in ghosts~\cite{de_felice_inevitable_2010,nojiri_ghost-free_2017}.
	A general discussion of such issues is beyond the scope of this paper.
	For the specific case we will be discussing, this will not be relevant.
	
	Cosmological effects of $f(R,T)$ theories have been explored by choosing several functional forms of $f$.
	The separation $f(R,T)=f_1(R)+f_2(T)$ has received much attention because one can explore the contributions from $T$ without specifying $f_1(R)$.  
	For example, in such separable theories, a non-equilibrium picture of thermodynamics at the apparent horizon of the Friedmann-Lema\^itre-Robertson-Walker (FLRW) universe was studied in~\cite{sharif_thermodynamics_2012}.
	
	While higher powers of $T$ have been considered~\cite{zaregonbadi_dark_2016}, at low densities the linear contributions will dominate, and linear $f_2(T) \propto T$ is of interest and has been studied \cite{harko_frt_2011,carvalho_stellar_2017,velten_cosmological_2017,deb_strange_2018}.
	
	Harko \textit{et al}.~\cite{harko_frt_2011} noted that such a simple model could produce cosmological constant like effects, but it is easy to see from their formula (29) that an accelerating universe is not possible from this term alone.
	As we will demonstrate, strong limits on $\chi$, which we use to parameterize linear contributions from $T$, indicate that this term has no cosmological significance.
	
	Even if $f(R,T)$ is not separable, one would expect that in situations of low curvature and matter density, the linear terms of $f(R,T)$ should dominate.  We therefore focus on these terms, and assuming there is no constant term (cosmological constant) we may approximate
	\begin{equation}\label{eq:fRT_linear}
		f(R,T) = R+2\chi T,
	\end{equation}
	where the coefficient of $R$ must be one to yield conventional gravity in low curvature environments, and $\chi$ is a single parameter describing the modification of gravity.
	Note there is no question of Ostrogradsky instability in this theory.
	Obtaining strong limits on $\chi$ could severely effect the possible contributions of this term in astrophysical situations.
	
	One approach to studying the consequences of such gravitational modifications would be comparison with limits from the parameterized post-Newtonian (PPN) formalism~\cite{will_confrontation_2014}.
	Strong limits on several parameters from the Solar System and other astrophysical systems can be obtained.
	Such an approach is non-trivial in this case because the dynamics of this theory are not entirely described in terms of the metric.
	In particular, as will be shown below, the stress-energy tensor is not conserved, and hence pressureless dust will not generally follow geodesics.
	We do not pursue this approach here, primarily because we believe we can achieve stronger limits by using the Tolman-Oppenheimer-Volkoff (TOV) equations.
	
	In~\cite{deb_strange_2018}, modified TOV equations were obtained and used to obtain modifications to models of strange stars.
	It was found that substantial and potentially measurable changes to such stars occurred if $|\chi| \sim 1$.  
	Observational data for white dwarfs were used in~\cite{carvalho_stellar_2017} to obtain a lower limit of $\chi \gtrsim -3\times 10^{-4}$.
	This limit was obtained by modeling the interior of the white dwarf as a non-interacting zero temperature electron gas.
	Indeed,~\cite{carvalho_stellar_2017} was unable to obtain self-consistent solutions for $\chi >0$, because the vanishing of the sound velocity near the surface of the white dwarf did not allow the density to drop to zero at finite radius.
	However, the surface of the white dwarf is not at zero temperature, and the electron interactions are not negligible, so we believe that positive $\chi$ values are also allowed.
	We have not performed this calculation, because we believe we can obtain more stringent limits by considering the Earth's atmosphere.
	
	In this paper, we investigate Eq.~\eqref{eq:fRT_linear} in the weak-field regime of Earth's atmosphere to further limit the possible range of $\chi$.
	By modeling the atmosphere as a perfect fluid ideal gas, we obtain modifications to the traditional hydrostatic equation
	\begin{equation}
		\frac{dp}{dr}=-g\left(p+\rho\right),
	\end{equation}
	where $p$ is isotropic pressure, $\rho$ is mass density, and $g$ is acceleration due to gravity.
	Solutions to the modified hydrostatic equation can then be compared to the atmospheric model given in~\cite{nasa_u.s._1976} to obtain strong limits on $\chi$.
	
	Our paper is structured as follows.
	The general formalism for $f(R,T)$ gravity is given in Section~\ref{sec:Formalism}.
	In Section~\ref{sec:HydrostaticEquation}, we derive the modified hydrostatic equation in a spherically symmetric, static spacetime.
	The hydrostatic equation is then obtained for our model of Earth's atmosphere and computational results and limits on $\chi$ are given in Section~\ref{sec:Earth'sAtmosphereAndResults}.
	Finally, our conclusions are given in Section~\ref{sec:Conclusion}.
	
	We use the sign conventions of Misner, Thorne, and Wheeler~\cite{misner_gravitation_1973} with metric signature $(-+++)$ and work in units where $c=G=1$.

\section{$\fRT$ formalism}\label{sec:Formalism}
	The theory of $f(R,T)$ gravity is motivated by the $f(R)$ framework that replaces the standard Hilbert action with an arbitrary function of the Ricci scalar $R$~\cite{nojiri_unified_2011}.
	Harko \textit{et al}.~\cite{harko_frt_2011} first proposed $f(R,T)$ gravity by introducing to the gravitational action an arbitrary dependence on the trace of the stress-energy tensor $T$.
	The full action is
	\begin{equation}\label{eq:FullAction}
		S = \int d^{4}x \rootg \left[ \frac{1}{16\pi} f(R,T) + \Lm \right],
	\end{equation}
	where the matter Lagrangian $\Lm$ describes any matter contributions.
	We will follow the derivation given by~\cite{harko_frt_2011}.\footnote{
		Ref.~\cite{harko_frt_2011} has an equation apparently identical to our Eq.~\eqref{eq:StressEnergyTensor}; however, they should have the opposite sign because they are working with the opposite sign metric.
		This apparent sign error is canceled by another apparent sign error when they choose $\Lm=-p$ as the perfect fluid matter Lagrangian.
	}
	Beginning with Eq.~\eqref{eq:FullAction}, we define the stress-energy tensor as
	\begin{equation}\label{eq:StressEnergyTensor}
		T_{\mu\nu} \equiv
		\frac{-2}{\rootg} \frac{\delta ( \rootg \Lm )}{ \delta g^{\mu\nu} }
		= -2 \frac{ \delta \Lm }{ \delta g^{\mu\nu} } + g_{\mu\nu} \Lm.
	\end{equation}
	There is an implicit assumption that $\Lm$ does not depend on derivatives of the metric.
	Variation of Eq.~\eqref{eq:FullAction} with respect to $g^{\mu\nu}$ yields the field equations
	\begin{multline}\label{eq:FieldEquations}
		( R_{\mu\nu} + g_{\mu\nu}\Box - \nabla_{\mu} \nabla_{\nu} ) \fr - \frac{1}{2} f(R,T) g_{\mu\nu}\\
		= -\ft ( \Theta_{\mu\nu} + T_{\mu\nu} ) + 8\pi T_{\mu\nu},
	\end{multline}
	where $\fr \equiv \partial f(R,T) / \partial R$, $\ft \equiv \partial f(R,T) / \partial T$, and 
	\begin{equation}\label{eq:Theta}
		\Theta_{\mu\nu} \equiv g^{\alpha\beta} \frac{\delta T_{\alpha\beta}}{\delta g^{\mu\nu}}
		= g_{\mu\nu} \Lm - 2 T_{\mu\nu} - 2 g^{\alpha\beta} \frac{\partial^2 \Lm}{\partial g^{\mu\nu} \partial g^{\alpha\beta}}.
	\end{equation}
	The covariant derivative of Eq.~\eqref{eq:FieldEquations} can then be written as
	\begin{multline}\label{eq:StressEnergyDivergence}
		\nabla_{\mu} T^{\mu\nu} =
		\frac{\ft}{8\pi - \ft} \bigg\{
		\nabla_{\mu} \left[ \ln \ft \right] ( T^{\mu\nu} + \Theta^{\mu\nu} ) \\
		+ \nabla_{\mu} \left( \Theta^{\mu\nu} - \frac{1}{2} g^{\mu\nu} T \right)
		\bigg\}.
	\end{multline}
	Note that the stress-energy tensor in traditional and $f(R)$ gravity is divergenceless.
	Applying our explicit form from Eq.~\eqref{eq:fRT_linear}, this simplifies to 
	\begin{equation}\label{eq:SpecificDivergence}
		\nabla_{\mu} T^{\mu\nu} = \frac{\chi}{4\pi - \chi} \nabla_{\mu} \left( \Theta^{\mu\nu} - \frac{1}{2} g^{\mu\nu} T \right).
	\end{equation}
		
\section{Hydrostatic equation in spherically symmetric $\fRT$ gravity}\label{sec:HydrostaticEquation}
	Static, spherically symmetric objects are described by the metric
	\begin{equation}
		ds^2 = -e^{\nu(r)} \,dt^2 + e^{\lambda(r)}\,dr^2 + r^2 \left( d\theta^2 + \sin^2\theta \,d\phi^2 \right),
	\end{equation}
	where $\nu(r)$ and $\lambda(r)$ are metric potentials.
	
	We will consider the stress-energy tensor of a perfect fluid, such that
	\begin{equation}
		T^{\mu\nu} = \left( p + \rho \right) u^{\mu} u^{\nu} + p g^{\mu\nu},
	\end{equation}
	where $\rho$ and $p$ are the energy density and isotropic pressure of the fluid, and $u^{\mu}$ is the fluid four-velocity, satisfying $u_{\mu} u^{\mu} = -1$ and $u^{\mu} \nabla_{\nu} u_{\mu} = 0$.
	Eq.~\eqref{eq:Theta} can now be written as
	\begin{equation}
		\Theta^{\mu\nu} = p g^{\mu\nu} - 2 T^{\mu\nu}.
	\end{equation}
	
	The field equations in traditional general relativity have no direct dependence on $\Lm$, leading to non-unique choices such as $\Lm=-\rho$ or $\Lm=p$, as discussed in~\cite{brown_action_1993}.
	In theories with non--minimal coupling of matter to curvature or an action with contributions from $T$, $\Lm$ explicitly appears in the field equations, and the choices for $\Lm$ become non-equivalent~\cite{bertolami_nonminimal_2008}.
	Following the work of~\cite{harko_frt_2011} and~\cite{carvalho_stellar_2017}, we will use $\Lm=p$.
	Then the hydrostatic equation from the radial component of Eq.~\eqref{eq:SpecificDivergence} is
	\begin{equation}\label{eq:HydrostaticEq}
		\frac{dp}{dr} = -g \left(\rho + p \right) + \frac{\chi}{8\pi + 2\chi} \frac{d}{dr} \left( \rho - p \right),
	\end{equation}
	where $g$ is the gravitational acceleration,
	\begin{equation}\label{eq:GravitationalAcceleration}
		g \equiv \frac{1}{2} \frac{d\nu}{dr}.
	\end{equation}

\section{Earth's atmosphere and results}\label{sec:Earth'sAtmosphereAndResults}
	In Earth's weak field, the atmosphere can be described using a Schwarzschild metric, for which $e^{\nu(r)}=1-2GM/r$ and thus $ g = GM/\left(r^2-2GMr\right) = g(r)$.
	In the atmosphere, $g(r) \approx GM/r^2$.
	We approximate $p\ll\rho$ and take Earth's atmosphere to be an ideal gas with $\rho = pM/RT$, where $M$ is atmospheric molar mass, $T$ is temperature, and $R$ is the ideal gas constant.
	After reinserting factors of $c$ and assuming constant $M$, Eq.~\eqref{eq:HydrostaticEq} is
	\begin{equation}\label{eq:IdealGasHydrodynamicEq}
	\frac{dp}{dr} = - \frac{gM}{RT}p + \frac{\chi M c^2}{\left(8\pi+2\chi\right) R} \frac{d}{dr}\left(\frac{p}{T}\right).
	\end{equation}
	
	We now introduce geopotential altitude $Z$, as defined in the U.S. Standard Atmosphere 1976~\cite{nasa_u.s._1976} as $g_0\,dZ = g\,dr$, where $g_{0} = 9.80665 \,\text{m}/s^2$ is the standard surface gravity and $Z=0$ corresponds to sea level.
	Hence, $Z$ will not exactly correspond to physical altitude; for example, a geopotential altitude of $Z=79\,\text{km}$ corresponds to a physical altitude of about $86\,\text{km}$ above sea level.
	Eq.~\eqref{eq:IdealGasHydrodynamicEq} in terms of geopotential altitude becomes
	\begin{equation} \label{eq:StAtmHydrodynamicEq}
	\frac{dp}{dZ} = -\frac{g_{0}M}{RT}p + \frac{\chi M c^2}{\left(8\pi+2\chi\right) R} \frac{d}{dZ}\left(\frac{p}{T}\right).
	\end{equation}
	
	\begin{figure}[t]
		\centering
		\includegraphics[width=\linewidth]{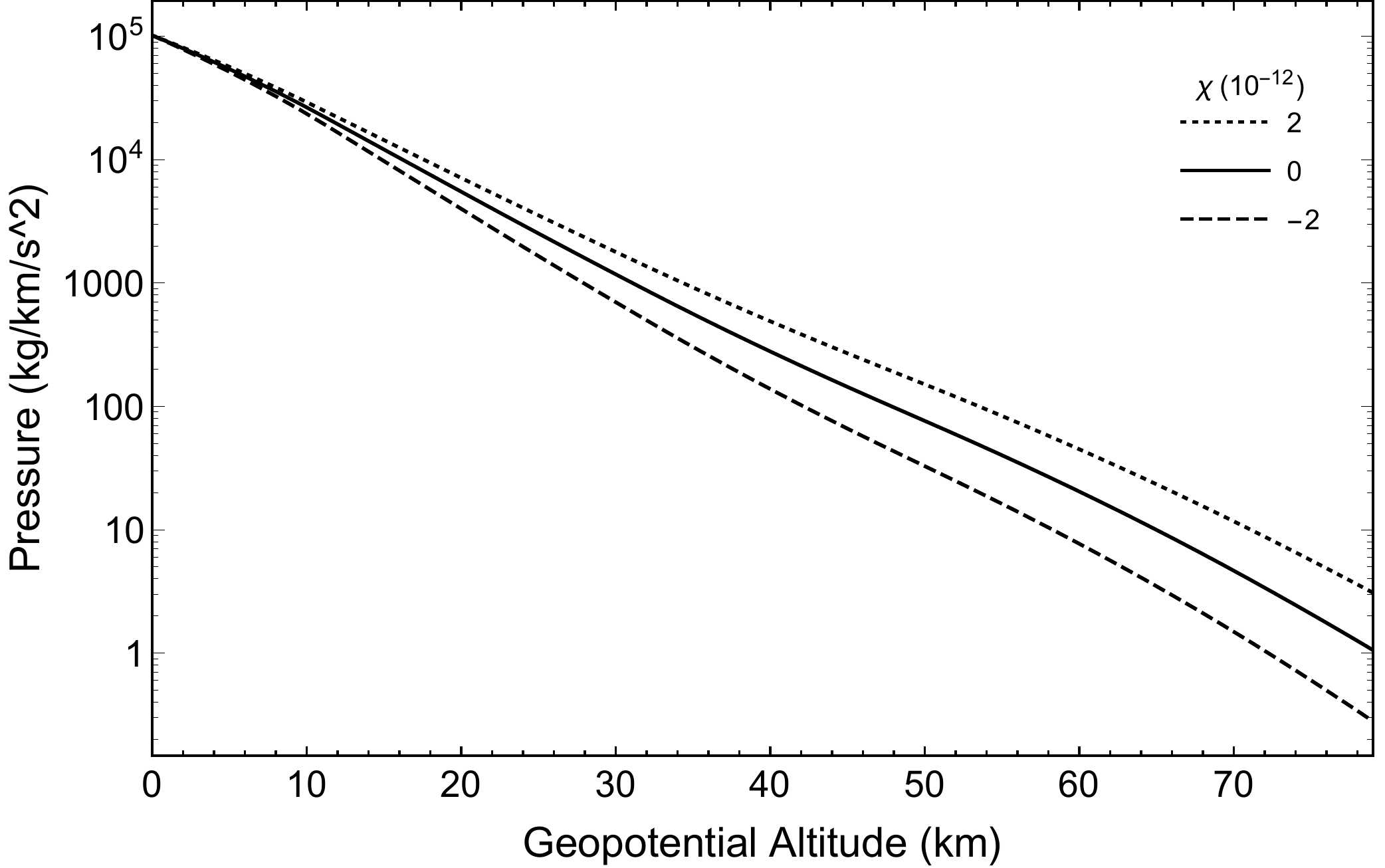}
		\caption{Pressure profile numerical solutions as a function of geopotential altitude $Z$.}
		\label{fig:Numerical}
	\end{figure}
	\begin{figure}[t]
		\centering
		\includegraphics[width=\linewidth]{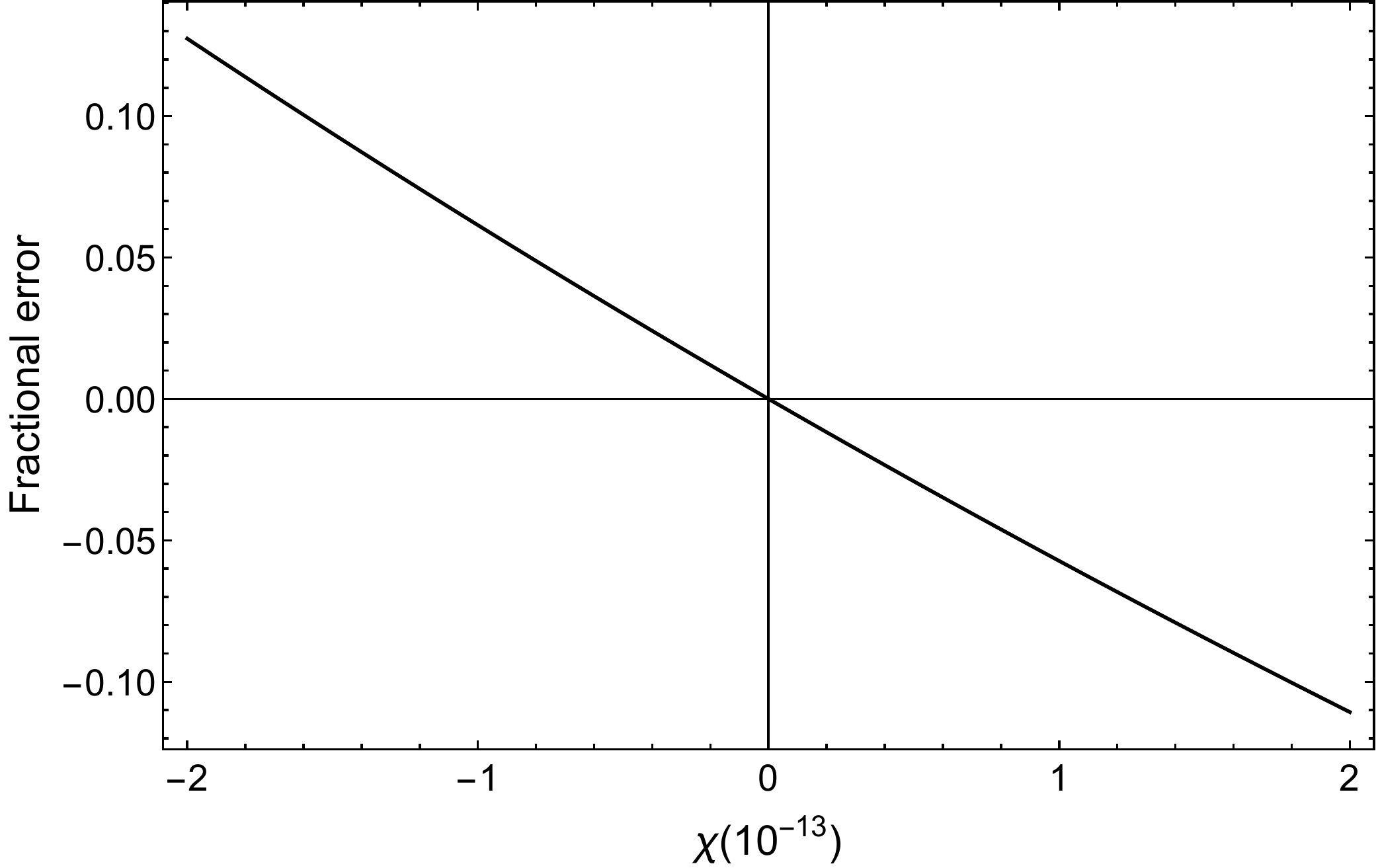}
		\caption{Fractional error of numerical solutions from U.S. Standard Atmosphere pressure model at $Z=79\,\text{km}$, which corresponds to a physical altitude of $86\,\text{km}$.}
		\label{fig:FractionalError}
	\end{figure}
	
	We use the U.S. Standard Atmosphere model for both temperature and atmospheric composition as a function of geopotential height.
	The U.S. Standard Atmosphere was an update to an existing international model first published in 1958 and updated in 1962, 1966, and 1976 that developed a mathematical model of the atmospheric profile~\cite{nasa_u.s._1976}.
	It proposed a pressure equation similar to Eq.~\eqref{eq:StAtmHydrodynamicEq} without $\chi$ terms.
	The model also developed geopotential profiles for mass density, temperature, and other atmospheric measurements.
	It separated Earth's atmosphere into upper and lower regions, with the boundary at $Z= 79\,\text{km}$.
	We focus on the lower region as it is the best measured and understood.
	In this region, composition is approximately constant, with $M=28.9644\times 10^{-3} \,\text{kg}/\text{mol}$, and the temperature is a piecewise linear function of $Z$.
	
	The final update to the U.S. Standard Atmosphere improved the model such that it differed from atmospheric measurements by at most $10\%$~\cite{nasa_u.s._1976}.
	We assert that any changes to the predicted pressure profile due to $\chi$ terms can be no larger than the largest error of $10\%$ from the model.
	We numerically solved the differential equation in Eq.~\eqref{eq:StAtmHydrodynamicEq} for pressure given specific values of $\chi$.
	Fig.~\ref{fig:Numerical} shows pressure profiles for select $\chi$ values as a function of $Z$, and Fig.~\ref{fig:FractionalError} shows the fractional change to the pressure as a function of $\chi$ at $Z=79\,\text{km}$, where the pressure profile difference due to $\chi$ is greatest.
	From the analysis, we obtain approximate limits on $\chi$ of
	\begin{equation}
		-1.6\times 10^{-13} \lesssim \chi \lesssim 1.8\times 10^{-13}.
	\end{equation}

\section{Conclusion}\label{sec:Conclusion}
	We have examined the $f(R,T)$ gravity modified hydrostatic equation in Earth's atmosphere to obtain limits on the leading order contributions from $T$ to the gravitational action.
	We found that the parameter $\chi$ describing the leading order modification to gravity is limited to the range $-1.6\times 10^{-13} \lesssim \chi \lesssim 1.8\times 10^{-13}$ by examining the atmospheric region below $Z=79\,\text{km}$ (approximately an altitude of $86\,\text{km}$).
	These results follow from the assertion that modifications from $\chi$ should vary from observational data by at most $10\%$, which is motivated by the U.S. Standard Atmosphere model having a maximum percent deviation of the same amount.
	
	Within the limits we observe, leading order $T$ terms in the gravitational action would have tiny effects on strange stars (as studied in~\cite{deb_strange_2018}) and cosmological models.
	For example, Ref.~\cite{velten_cosmological_2017} found significant cosmological effects for $\chi\sim 1$.
	Our limits are also nine orders of magnitude stronger than limits from white dwarfs~\cite{carvalho_stellar_2017}.
	
	Perhaps more promising would be to look at other terms whose contributions would be small in Earth's atmosphere but could be larger in more extreme situations.
	The term $\gamma R T$, for instance, would yield an extra $R$ term in the covariant derivative of the stress-energy tensor, which in environments of large curvature, such as neutron stars, could produce significant changes to traditional models.
	
	\begin{acknowledgments}
		We would like to thank P.~Anderson for his helpful discussion.
	\end{acknowledgments}

\end{document}